\documentclass[aps,prb,twocolumn,superscriptaddress]{revtex4}

\usepackage{graphicx}  
\usepackage{dcolumn}   
\usepackage{bm}        
\usepackage{amssymb}   
\usepackage{amsmath}

\begin{document}


\title{Electron-phonon scattering from Green's function transport combined with Molecular Dynamics: Applications to mobility predictions.}


\author{Troels Markussen}
\email[]{troels.markussen@quantumwise.com}
\affiliation{QuantumWise A/S, Fruebjergvej 3, Postbox 4, DK-2100 Copenhagen, Denmark}
\author{Mattias Palsgaard}
\affiliation{QuantumWise A/S, Fruebjergvej 3, Postbox 4, DK-2100 Copenhagen, Denmark}
\affiliation{Department of Micro- and Nanotechnology (DTU Nanotech), Center for Nanostructured Graphene (CNG), Technical University of Denmark, DK-2800 Kgs. Lyngby, Denmark}
\author{Daniele Stradi}
\affiliation{QuantumWise A/S, Fruebjergvej 3, Postbox 4, DK-2100 Copenhagen, Denmark}
\author{Tue Gunst}
\affiliation{Department of Micro- and Nanotechnology (DTU Nanotech), Center for Nanostructured Graphene (CNG), Technical University of Denmark, DK-2800 Kgs. Lyngby, Denmark}
\author{Mads Brandbyge}
\affiliation{Department of Micro- and Nanotechnology (DTU Nanotech), Center for Nanostructured Graphene (CNG), Technical University of Denmark, DK-2800 Kgs. Lyngby, Denmark}
\author{Kurt Stokbro}
\affiliation{QuantumWise A/S, Fruebjergvej 3, Postbox 4, DK-2100 Copenhagen, Denmark}



\date{\today}

\begin{abstract}
We present a conceptually simple method for treating electron-phonon scattering and phonon limited mobilities. By combining Green's function based transport calculations and molecular dynamics (MD), we obtain a temperature dependent transmission from which we evaluate the mobility. We validate our approach by comparing to mobilities and conductivies obtained by the Boltzmann transport equation (BTE) for different bulk and one-dimensional systems. For bulk silicon and gold we successfully compare against experimental values. We discuss limitations and advantages of each of the computational approaches.
\end{abstract}


\maketitle

\section{Introduction \label{intro}}
The continued down-scaling of electronic devices and interconnects calls for accurate simulation models which incorporates the effects of quantum confinement of both electrons and phonons, surface effects, strain etc. It is increasingly difficult to describe all such effects with continuum models, which are typically parametrized to fit bulk materials. Atomistic models, on the other hand can describe many of the important effects. Density functional theory (DFT) is particularly important in this respect since it is a first-principles method which doesn't need to be fitted to a particular device, while it is computationally possible to study systems with several thousands of atoms.

Electron-phonon coupling (EPC) plays a central role in the performance of most electronic devices. Several recent studies have studied EPC in bulk materials by calculating the EPC from first principles and using the Boltzmann transport equation (BTE) for evaluating the electron mobility and conductivity\cite{Hwang2008a,Chen2009,Kaasbjerg2012,Borysenko2010,Restrepo2014,Park2014,Yan2009,Gunst2016}. Bulk calculations of EPC can, however be rather demanding as one need to integrate the coupling over both electron and phonon wavevectors ($\mathbf{k}$- and $\mathbf{q}$-space). Moreover, EPC in amorphous materials can only be calculated approximately with the BTE approach.

Atomistic modelling of electronic devices is typically carried out using non-equilibrium Green's function (NEGF) theory in combination with either DFT\cite{Brandbyge2002} or tight-binding methods\cite{LuisierPRB2009, ATK-SE}. EPC can be rigorously included in NEGF using perturbation theory. However, the resulting equations are numerically very challenging and approximations need to be applied. Approximate methods include lowest order expansions of the inelastic current\cite{FrederiksenPRB2007,JingTaoPRB2014-XLOE} or approximations to the EPC self-energy\cite{LuisierPRB2009, LuisierPRB2014}.

With few exceptions\cite{LuisierAPL2014}, most of EPC with both the BTE and within NEGF assumes that the phonons can be described within the harmonic approximation, since addition of anharmonic effects significantly increases the computational burden. However, at room temperatures and above there will be anharmonic contributions to the phonons for many materials.
Molecular dynamics (MD) simulations, on the other hand, inherently includes anharmonic effects without extra computational requirements.

Previous works have used MD simulations in combination with standard Landauer transmission calculations. This has primarily been done in order to sample different configurations of e.g. a molecule in contact with two metallic electrodes\cite{Li2004,Andrews2008,Paulsson2008,SolomonJACS2008}, metallic point contacts\cite{Brandbyge1997, Dreher2005}, and carbon nanotubes\cite{Pecchia2003-CNT}. A single study used MD simulations to actually probe the energy dependent EPC\cite{Pecchia2003}. MD simulations have also been successfully applied in combination with DFT calculations to calculate the temperature dependent band structure of bulk and nanocrystals of silicon\cite{Franceschetti}. Recently, Liu et al. \cite{KellyPRB2015} used a very similar approach to successfully obtain mean free paths and resistivities in bulk metals.

In this work, we develop a MD-Landauer approach for calculating the temperature dependent mobility and conductivity. In a device geometry with a central region coupled to two electrodes, we perform MD simulations in a certain part of the central region. When increasing the length of the MD region we obtain a length dependent resistance. From the slope of the linearly increasing resistance vs. length curve we obtain the resistivity and eventually the mobility. Further computational details will be provided below.

We apply the method to various systems covering metals and semi-conductors as well as one, two, and three-dimensional systems. In order to validate our approach we compare the mobilities obtained from the MD-Landauer approach with results obtained from the BTE for the same systems. In general we find that the two approaches give similar results. Our BTE approach has been described in a previous work\cite{Gunst2016}. Here we further validate the BTE method by successfully comparing the obtained temperature dependent electron mobilities of bulk silicon with experimental values. All calculations have been performed with the Atomistix ToolKit (ATK) \cite{ATK}.

The paper is organized as follows: In Sec.~\ref{sec:methods} we first describe general details of our computational methods. The MD-Landauer method is detailed in Sec.~\ref{sec:MD-Landauer} where we also show results for a silicon nanowire. In Sec.~\ref{sec:bulk-silicon} we present results for bulk silicon and compare against experimental mobilities, while results for a gold nanowire and bulk gold is presented in Secs.~\ref{sec:Au-NW} and \ref{sec:Au-bulk}.
We discuss advantages and weaknesses of the two methods in Sec.~\ref{sec:discussion} and summarize our findings in Sec.~\ref{sec:conclusion}.

%

\section{Methods and Results \label{sec:methods}}
All results presented in this paper are obtained with ATK\cite{ATK}. The electronic structure, band energies, Hamiltonians and derivative of Hamiltonians are calculated from density functional theory (DFT) within the local density approximation (LDA) to the exchange-correlation functional.\cite{PerdewZunger1981} For the gold nanowires, we additionally compare properties obtained by density functional tight binding\cite{hotbit}.

Phonon energies and polarization vectors have been calculated from either DFT or from classical force field potentials (details will be provided for each studied system). In the case of MD simulations we only use the classical potentials.

Our implementation of the electron-phonon coupling and BTE has been documented and verified in Ref.~\onlinecite{Gunst2016} for various two-dimensional systems.

\subsection{MD-Landauer approach \label{sec:MD-Landauer}}
In this section we present the details in our MD-Landauer approach for treating electron-phonon scattering. We will illustrate the method by showing calculations for the 1.5 nm [110] silicon nanowire (SiNW). A similar procedure is used for the other systems presented below.

\begin{figure}[htb!]
\includegraphics[width=0.9\columnwidth]{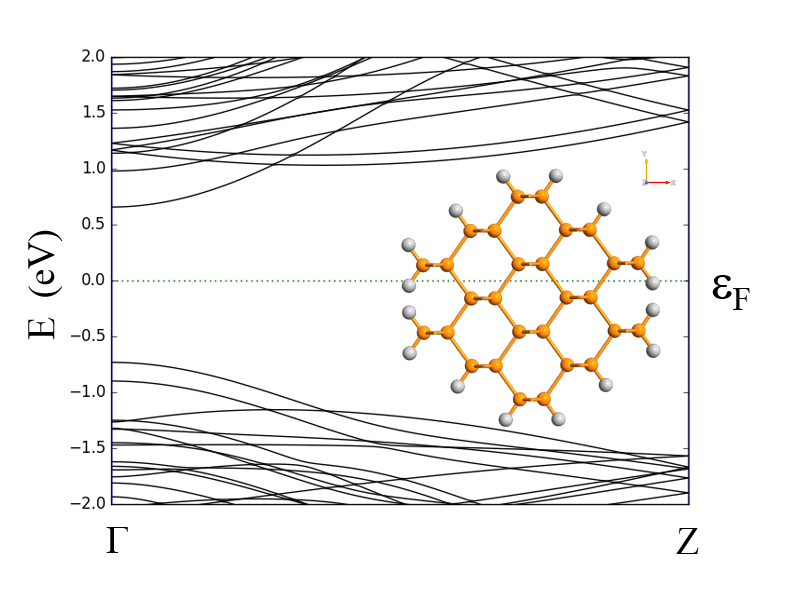}
\caption{Electronic band structure of the SiNW. The inset shows a cross sectional view of the SiNW. \label{fig:SiNW-band-and-struct}}
\end{figure}

Figure \ref{fig:SiNW-band-and-struct} shows the electronic band structure for the SiNW. The conduction band minimum is 0.66~eV above the Fermi energy (at 0.0~eV in the plot). The next conduction band is 0.32~eV higher in energy and for electron transport close to the conduction band minimum (CBM) it is sufficient to include the lowest conduction band.

\begin{figure}[htb!]
\includegraphics[width=0.9\columnwidth]{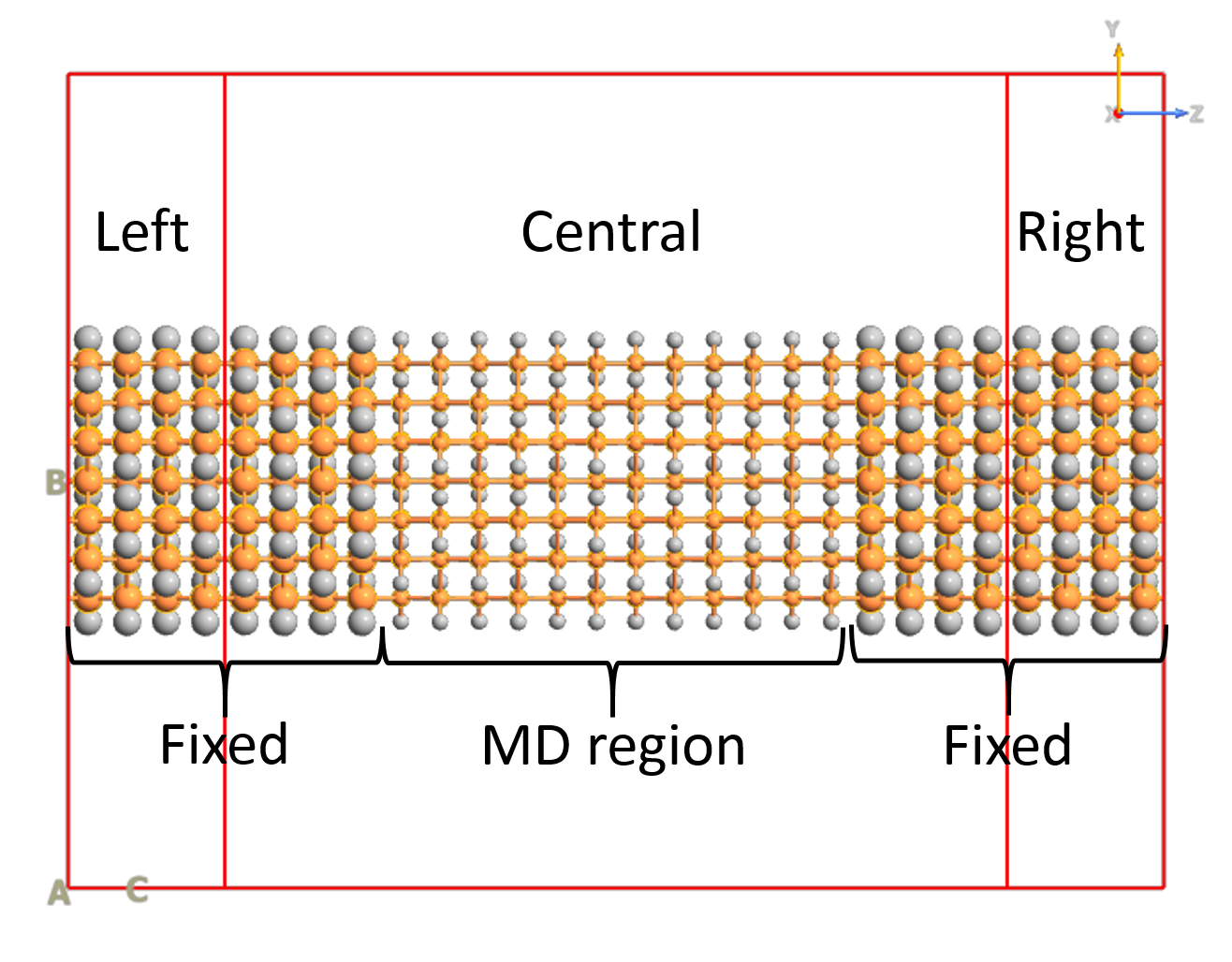}
\caption{Device setup for the MD-Landauer approach. A central region is coupled to two semi-infinite electrodes (Left and Right). Initially the wire is periodic in the z-direction. An MD simulation is performed for the atoms in the middle part of the central region (MD region), while the electrodes as well as the first electrode copy on either side of the central region are kept fixed at their equilibrium positions. The fixed atoms are drawn as larger spheres.\label{fig:md-landauer-setup}}
\end{figure}

We now consider a device configuration where a central region is coupled to two semi-infinite electrodes (Left and Right) as shown in Fig. \ref{fig:md-landauer-setup}. The atomic positions in the electrodes are kept at their equilibrium positions and so are the first copy of the electrodes inside the central region. In the middle part of the central region, called \textit{MD region} in Fig. \ref{fig:md-landauer-setup}, the atoms are allowed to move according to a molecular dynamics simulation. For all the calculations presented in this paper, the MD simulations are performed using classical potentials which make the calculations very efficient. For the SiNW and bulk Si MD calculations we apply a Tersoff potential\cite{Tersoff2005}. For both the SiNW and Au nanowires we have compared the results from BTE using phonons calculated with DFT with phonons from classical calculations and found the obtained results qualitatively agree, thus justifying the use of classical potentials. We note that the MD approach does not displace atoms along a single phonon mode at a time. Hereby, the predictability of the MD-Landauer approach does not rely on the accurate description of a single phonon mode but rather the full configuration space including anharmonic effects.

\begin{figure*}[htb!]
\includegraphics[width=2\columnwidth]{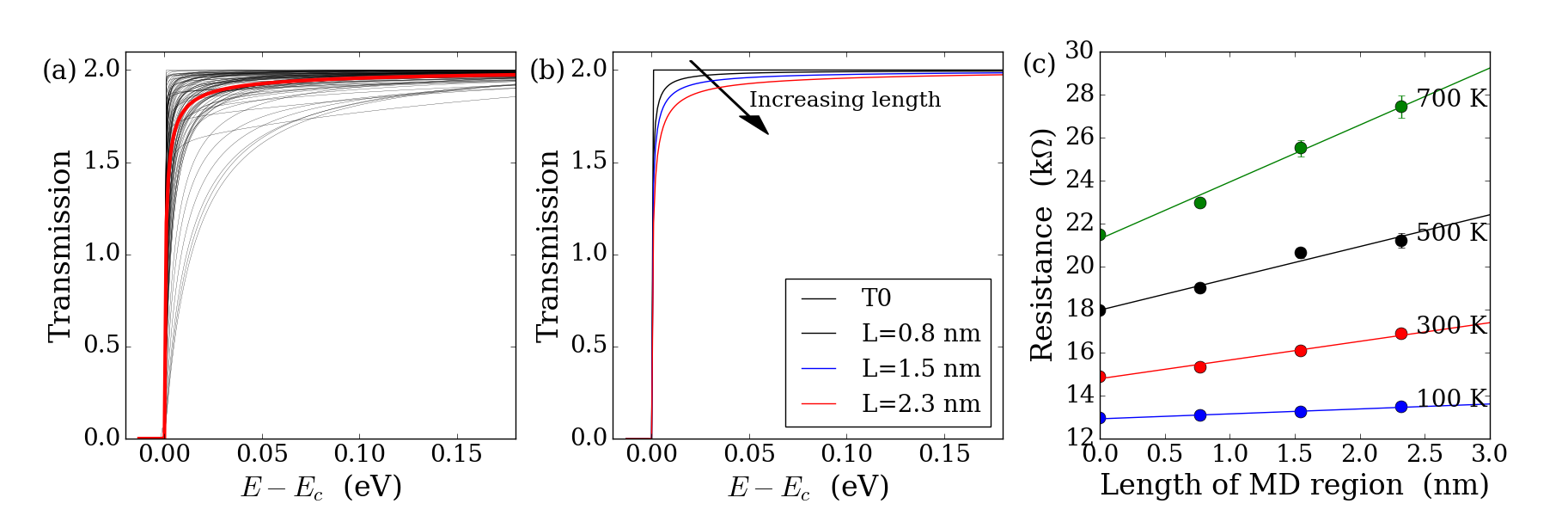}
\caption{(a) Transmission functions from different MD simulations (thin black lines) together with the average transmission (thick red). The length of the MD region is 2.3 nm and the temperature is 100 K. The average transmission functions at three different lengths are shown in panel (b). Panel (c) shows the resistance vs. length of the MD region for the SiNW at 100 K. \label{fig:sinw-transmissions}}
\end{figure*}


When the MD simulation is equilibrated we take a snapshot of the atomic configuration and calculate the electronic transmission function using DFT. This step first involves a self-consistent DFT-NEGF calculation of the device configuration and subsequently a calculation of the Landauer transmission function using a standard Green's function method:
\begin{equation}
\mathcal{T}(E,\mathbf{x}\{T\}) = {\rm Tr}[\mathbf{G}^r(E,\mathbf{x}\{T\})\mathbf{\Gamma}_L(E)\mathbf{G}^a(E,\mathbf{x}\{T\})\mathbf{\Gamma}_R(E)],
\end{equation}
where $\mathbf{G}^r=(E\mathbf{S}-\mathbf{H}(\mathbf{x}\{T\})-\mathbf{\Sigma}_L(E)-\mathbf{\Sigma}_R(E))^{-1}$ is the retarded Green's function in the central region described by the Hamiltonian $\mathbf{H}(\mathbf{x}\{T\})$ and overlap $\mathbf{S}(\mathbf{x}\{T\})$ matrices. The Hamiltonian and overlap matrices depend explicitly on the random displacement of the atoms ($\mathbf{x}$), which in turn depend on the temperature used in the MD calculation, as well as on the random initial velocities used in the MD simulations. Due to this randomness, we need to perform several MD simulations in order to  obtain a good sample averaging. The coupling to the semi-infinite left- and right electrodes are taken into account through the self-energies $\mathbf{\Sigma}_{L,R}(E)$, whose imaginary part give the coupling matrices $\Gamma_{L,R}(E) = -2{\rm Im}[\mathbf{\Sigma}_L(E)]$. All transmission functions are calculated at zero bias. Figure \ref{fig:sinw-transmissions} (a) shows the results of 100 individual MD + Landauer transmissions (thin black lines) as well as the average transmission function (thick red). Each MD calculation is started with a Maxwell-Boltzmann distribution of the velocities corresponding to the target temperature. We use a Langevin thermostat\cite{LangevinMD} with a time step of 1 fs. After an equilibration time of 5 ps we take a snapshot of the configuration and calculate the electronic transmission spectrum with DFT-NEGF. Since the MD calculations are very fast, we simply restart the MD calculations for each sample. Due to the random initial velocities, each MD simulation will result in different configurations after the same equilibration time.

This procedure is repeated at different lengths of the MD region, $\mathcal{L}$, with average transmissions shown in Fig. \ref{fig:sinw-transmissions} (b). From the average transmission $\langle \mathcal{T}_{\mathcal{L}}(E,T) \rangle$ we obtain the length- and temperature dependent conductance from the Landauer formula
\begin{eqnarray}
G(\mathcal{L},T) = \frac{2e^2}{h}\int \langle \mathcal{T}_{\mathcal{L}}(E,T)\rangle \left(-\frac{\partial f(E,\mu,T)}{\partial E}\right)dE  \label{eq:conductance},
\end{eqnarray}
where $f(E,\mu,T) = (e^{(E-\mu)/k_BT}+1)^{-1}$ is the Fermi-Dirac distribution function at chemical potential $\mu$. We allow ourself to freely adjust the chemical potential without explicitly taking doping effects into account. Notice that the average transmission $\langle \mathcal{T}_{\mathcal{L}}(E,T) \rangle$ only depend on the energy and temperature, when averaged properly. The remaining effect of randomness is represented by error bars in the plots of resistance and mobility presented below.

Figure \ref{fig:sinw-transmissions} (c) shows the resistance $R(\mathcal{L},T) = 1/G(\mathcal{L},T)$ vs. length of the MD region. The points show the average resistance, and the error bars show the standard deviations of the average resistance.
We observe that the resistance increases linearly with length showing that the resistance is ohmic. The linear fit to the averaged data is written as
\begin{eqnarray}
R(\mathcal{L},T) = R_c + \rho_{1D}(T) \mathcal{L},
\end{eqnarray}
thus defining a one-dimensional resistivity, $\rho_{1D}(T)$, which depends on temperature, but not on wire length. $R_c$ is the length independent contact resistance. Note that the one-dimensional resistivity has units of $\Omega/m$ whereas a usual bulk resistivity is measured in units of $\Omega\cdot m$. In order to convert the one-dimensional conductivity to a bulk quantity we must multiply with the wire cross sectional area $A$, i.e. $\rho_{bulk} = \rho_{1D}\cdot A$.

In order to calculate the mobility we need to determine the carrier density. From a separate calculation of the density of states, $D(E)$, of the bulk wire (evaluated at the equilibrium atomic structure) we calculate the carrier density per unit area

\begin{equation}
\tilde{n}= \frac{n}{A} = \int_{E_g}^\infty f(E,E_F,T) D(E)dE,
\end{equation}
where we use the middle of the band gap, $E_g$, as the lower integration limit in the case of electron mobilities. In the case of holes, we should integrate from $-\infty$ to $E_g$ and replace $f\rightarrow 1-f$.  We finally calculate the mobility as
\begin{equation}
\mu = \frac{1}{q\,n\,\rho_{bulk}} = \frac{1}{q\,\tilde{n}\,\rho_{1D}}.
\end{equation}
Notice that the final expression for the mobility does not depend on the wire cross sectional area.

\begin{figure}[htb!]
\includegraphics[width=\columnwidth]{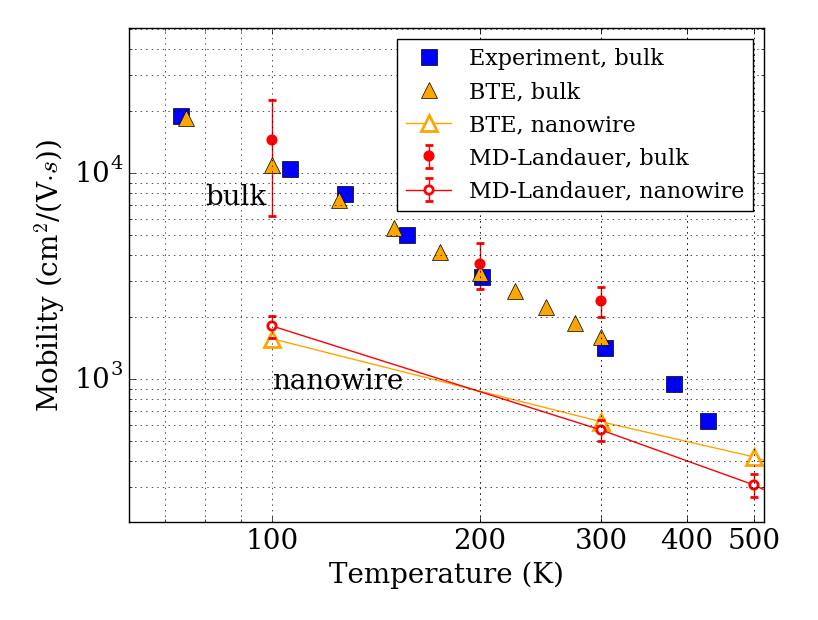}
\caption{Phonon limited mobility vs. temperature for the SiNW (open markers with lines) and bulk silicon (filled markers) calculated with BTE and with the MD-Landauer approach. The error bars for the MD-Landauer results indicate the standard deviations on the calculated mobilities. Experimental values for bulk silicon\cite{CanaliPRB1975} are shown for comparison (blue dots). \label{fig:SiNW-mobility-vs-temp.}}
\end{figure}

The resulting mobilities obtained with the MD-Landauer method are shown in Fig. \ref{fig:SiNW-mobility-vs-temp.} together with results obtained from the BTE. In the same figure we also show electron mobilities for bulk silicon obtained from BTE and MD-Landauer together with experimental values. 

For the BTE calculations electrons are calculated using DFT with local density approximation (LDA) exchange-correlation functional and double-zeta-polarized (DZP) basis set. 
Hamiltonian derivatives are calculated using supercells repeated 11 times along the [110]-direction.
The phonons are calculated using classical potentials.
When calculating the EPC we include only the lowest conduction band and use 150 k-points in the range [0,0.05]$\pi/a$, with $a$ being the unit cell length of the nanowire. All phonons are included and we use 100 q-points the range [-0.2,0.2]$\pi/a$. When calculating the mobility an energy broadening of 3 meV is used for the approximate delta function in the Fermi's golden rule expression of the phonon mediated transition rate between states.

The Fermi energy is shifted to 50~meV above the CBM corresponding to a doping of $9\cdot10^{19}$cm$^{-3}$. We first notice that the two computational methods give almost the same temperature dependent mobility. Second, we observe that the mobility of the SiNW is almost an order of magnitude smaller than the bulk values. More details about the bulk silicon calculations are presented below. The reduction of the mobility in nanowires is in good agreement with previous theoretical studies based on tight-binding models\cite{NiquetPRB2010}.

The reduced mobility in the nanowire can be traced back to the increased EPC in nanowires due to (i) reduced complexity in fulfilling the selection rules for energy and momentum matching due to band folding, and (ii) localization and mixing of corresponding bulk phonon modes. Scattering from surface modes is found to be insignificant.
The origin of the scattering is directly available from the BTE through the scattering rate with individual phonon modes (not shown). However, in the MD-Landauer approach part of this information is lost. On the other hand, it does not require one to store the scattering rate for all k, q and phonon mode indices. Being more memory efficient the MD-Landauer approach may therefore be more appealing as a design tool for complex systems with many degrees of freedom and for bulk systems that can be very memory demanding due to the large number of k, q and phonon mode combinations needed.

\subsection{Bulk silicon \label{sec:bulk-silicon}}
We next consider the phonon limited electron mobility in bulk silicon. 

\begin{figure}[htb!]
\includegraphics[width=\columnwidth]{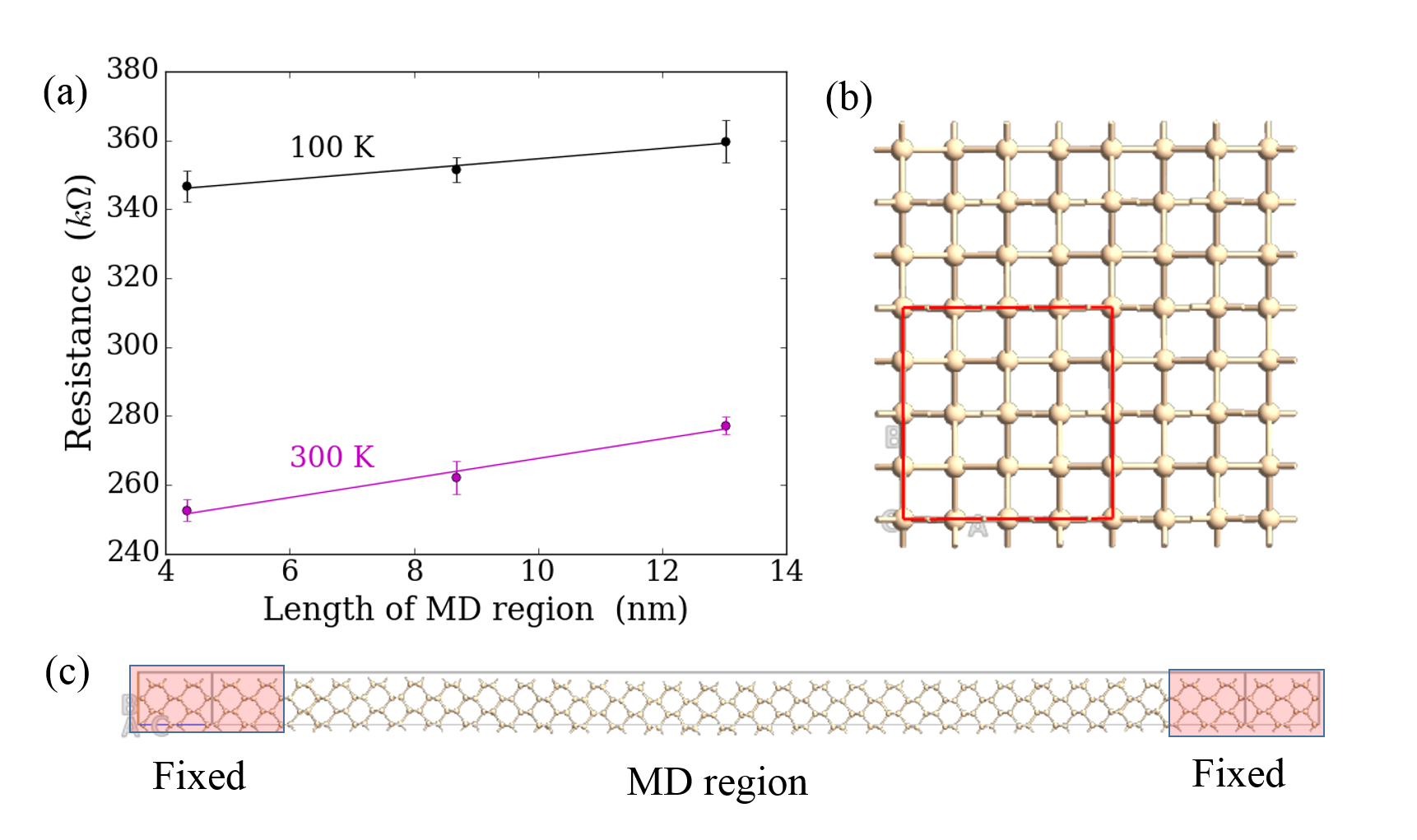}
\caption{(a) Length dependend resistance of bulk silicon at temperatures 100 K and 300 K. Panel (b) shows the cross section of the calculation cell (red box) while the device configuration is shown in (c). The length of the MD region is 13 nm.\label{fig:bulk-silicon-R-vs-L}}
\end{figure}

We performed MD-Landauer calculation for bulk silicon at temperatures 100 K, 200 K, and 300 K. The bulk silicon MD-Landauer calculations were performed on a $2\times2$ supercell with the transport in the [100] direction, as shown in Fig. \ref{fig:bulk-silicon-R-vs-L} (b) and (c), with the length of the MD region varying from 4 to 13 nm. We performed 20 different MD simulations in order to get averaged transmissions. The bulk silicon device has periodic boundary conditions in the $x$- and $y$ directions. For the self-consistent calculations we use an $11\times11$ transverse k-point sampling, while the transmission spectra were averaged with a $21\times21$ transverse k-point sampling. Figure \ref{fig:bulk-silicon-R-vs-L} (a) shows average resistances vs. length of MD region for temperatures 100 K (top curve) and 300 K (bottom curve). From the slope of the linear fits, we obtain the mobility as explained above. The error bars indicate the standard deviations on the average resistances.

For the BTE calculations, both phonons and electrons are calculated using DFT. The dynamical matrix and hamiltonian derivatives are calculated from super-cells constructed as a (7,7,7)-repetition of the primitive silicon unit cell (686 atoms). For the electron-phonon and mobility calculations we only consider a single conduction band valley and sample the electronic Brillouin zone (BZ) in local region around that valley with a sampling corresponding to a $99\times99\times99$ Monkhorst-Pack mesh. The phonon BZ is sampled with a $25\times25\times25$ mesh in a region around the $\Gamma$-point, with $|q|<0.075\pi/a$ with $a$ being the silicon lattice constant. All six phonon modes are included in the calculation.

In Fig. \ref{fig:SiNW-mobility-vs-temp.} we show the calculated electron mobility vs. temperature together with experimental data\cite{CanaliPRB1975}. It is evident that the BTE calculations reproduce the experimental data very well over the whole temperature interval. In addition to the BTE and experimental results, we also show the MD-Landauer results at temperatures 100 K, 200 K, and 300 K. It is encouraging that the MD-Landauer method gives mobility values in close agreement the the BTE- and experimental results.

\subsection{Au nanowire\label{sec:Au-NW}}
\begin{figure}[htb!]
\includegraphics[width=0.48\columnwidth]{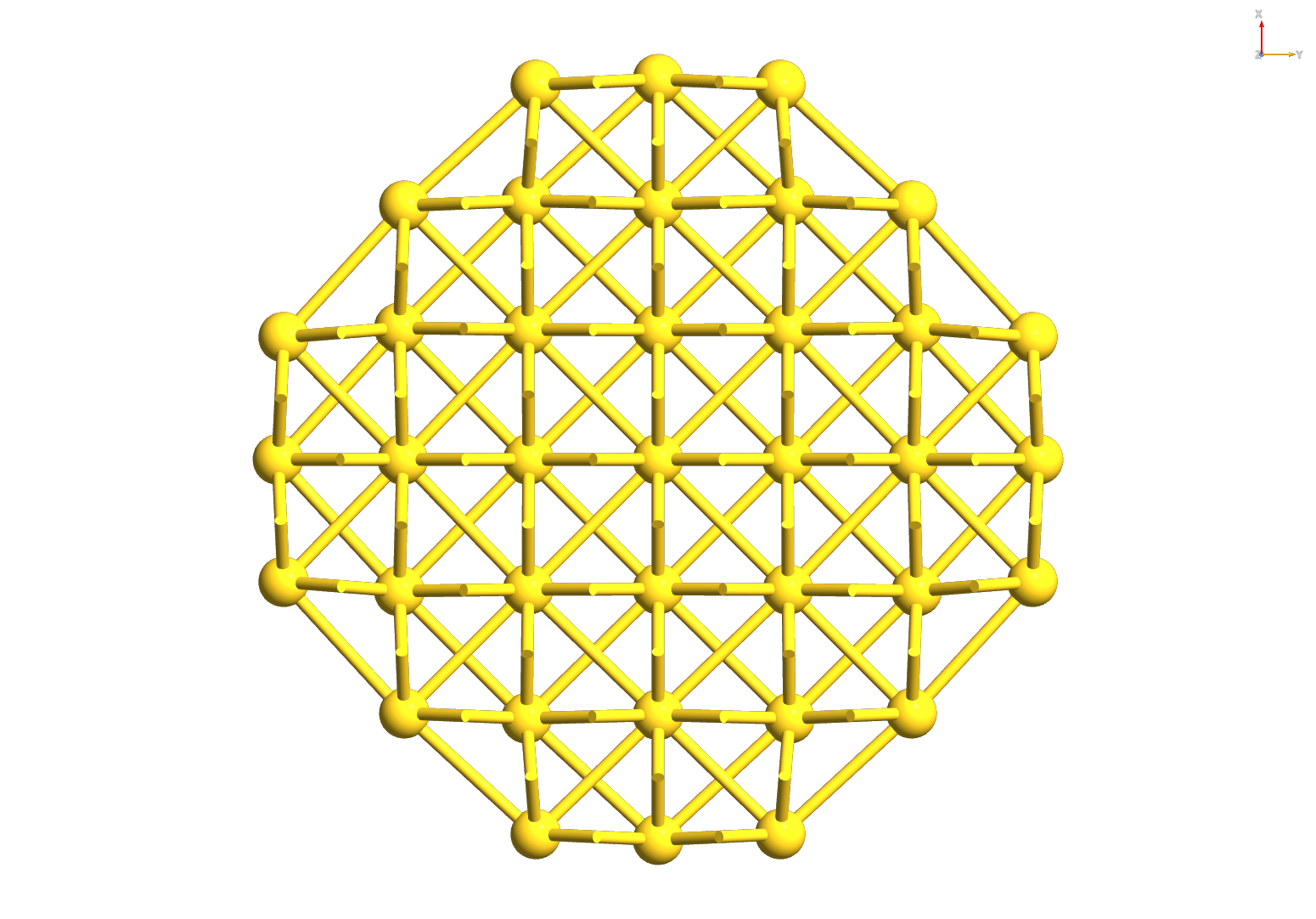}
\includegraphics[width=0.48\columnwidth]{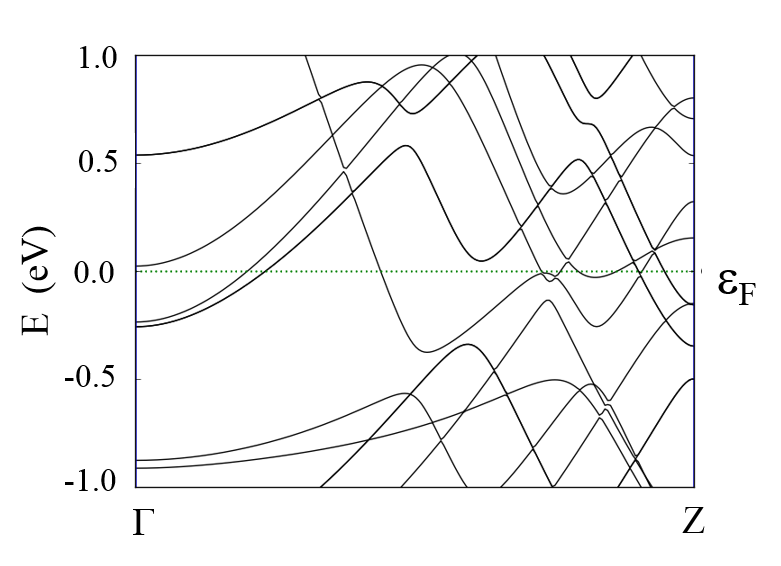}
\caption{Cross section of the Au NW (left) and electronic band structure (right). The diameter of the wire is approximately 1 nm. \label{fig:Au-cross-section-and-band}}
\end{figure}

We now continue to study metallic systems. The first system we consider is a thin gold nanowire with a diameter of 1.3 nm. A cross sectional view is shown in Fig. \ref{fig:Au-cross-section-and-band} (left) together with the electronic bandstructure calculated with a single-zeta-polarized basis set.

For the gold NW we have in addition to DFT performed calculations with a density functional tight binding (DFTB) description of the electrons.\cite{hotbit} For the BTE we have used phonons calculated from either DFT or from the embedded atom model (EAM)\cite{Sheng2011-EAM}. The MD simulations are only performed with EAM. In Table \ref{tab:Conductivities} we compare the nanowire resistivities obtained with the various methods (BTE or MD-Landauer) and how the parameters are calculated. The top row show results from BTE with both electrons and phonons obtained with DFT. The following rows show results where the phonons are described with EAM and the electrons either with DFT or DFTB. There is an overall good agreement between the two methods and the different parameters. It is computationally much more expensive to calculate the phonons from DFT than with the classical EAM. It is thus encouraging to see that the phonons seem to be accurately enough described with the EAM. This is true even though the EAM phonon energies are up to 30 \% lower in energy than the DFT phonon energies. 

In accordance with the silicon results, we find that the BTE and MD-Landauer methods give similar results for the resistivity, within a factor of two difference.

\begin{table}[htb!]
	\centering
		\begin{tabular}{llc}
		Method & Parameters & $\rho$ ($10^{-8} \times \Omega\cdot$m)  \\ \hline
		BTE & (DFT) & 5.6   \\
		BTE & (EAM+DFT) & 4.6   \\
    	BTE & (EAM+DFTB) & 3.8   \\
		MD-Landauer & (EAM+DFT) & 7.1   \\
		MD-Landauer & (EAM+DFTB) & 7.5
		\end{tabular}
	\caption{Resistivities of the Au nanowire at 300 K calculated in different ways. For reference, the experimental resistivity of bulk gold is 2.44$\times 10^{-8}\Omega\cdot$m at room temperature.}
	\label{tab:Conductivities}
\end{table}

\subsection{Au bulk\label{sec:Au-bulk}}
We now continue to study bulk gold in order to illustrate that our MD-Landauer approach also can be applied to bulk metallic systems, which were also studied in Ref. \onlinecite{KellyPRB2015} with a similar approach. For bulk calculations we set up a device with a cross section of $0.82\times0.82$nm$^2$ corresponding to a $5\times5$ repetition of the Au unit cell, when the transport is along the $[001]$ direction. We have verified that using $3\times3$ and $4\times4$ repetitions give essentially the same results. When calculating the transmission through the bulk system, we average over $6\times6$ transverse $\mathbf{k}$-points. The MD simulations are performed with the EAM and the electronic structure and transmission function is calculated with DFTB. We have verified for a single temperature (300 K) that calculating the electronic properties with DFT give essentially the same resistivity.

\begin{figure}[htb!]
\includegraphics[width=\columnwidth]{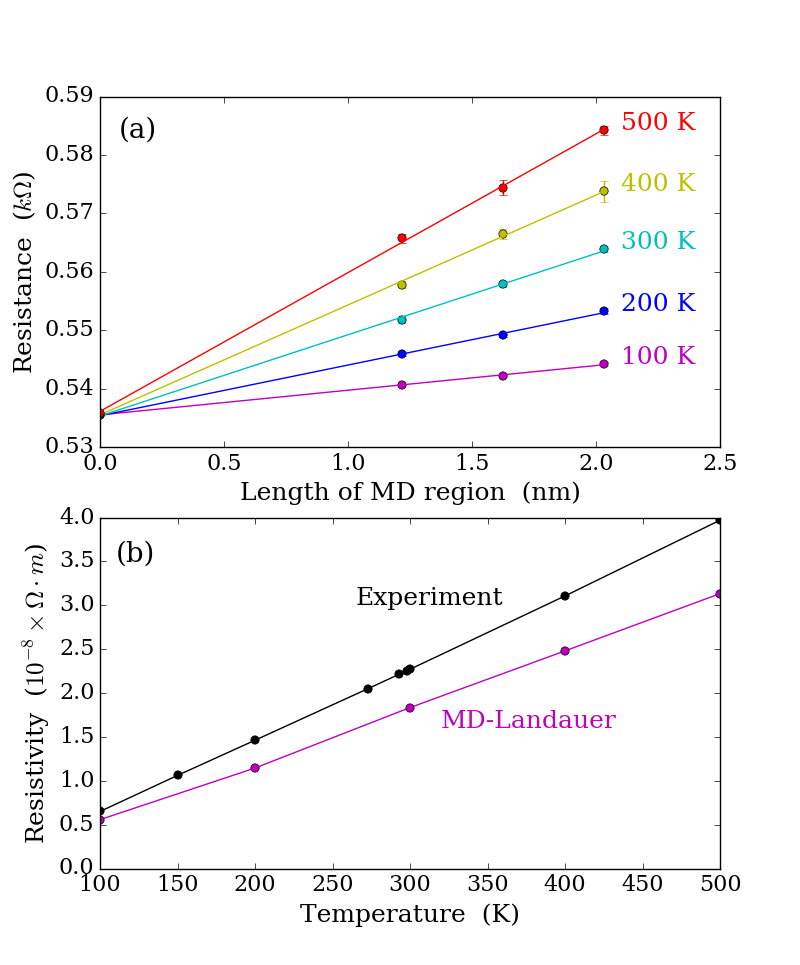}
\caption{Length dependend resistance at different temperatures (a) and temperature dependent resistivity (b). The resistivity is calculated for bulk Au with the MD-Landauer method. All results in this figure are obtained with EAM for the phonons/MD and DFTB for the electronic parts. The black points show experimental results\cite{HandbookOfPhysics}.\label{fig:Au-bulk-results}}
\end{figure}

Figure \ref{fig:Au-bulk-results} (a) shows the resistance vs. length of MD region for increasing temperatures. At all temperatures the resistance increases linearly with length and the resistivity is thus well defined. Panel (b) show the calculated temperature dependent resistivity of bulk gold together with experimental values\cite{HandbookOfPhysics}. The MD-Landauer method give bulk gold resistivities which are in very good agreement with the experimental results with about 20 \% difference.

\section{Discussion \label{sec:discussion}}
\begin{figure}[htb!]
\includegraphics[width=\columnwidth]{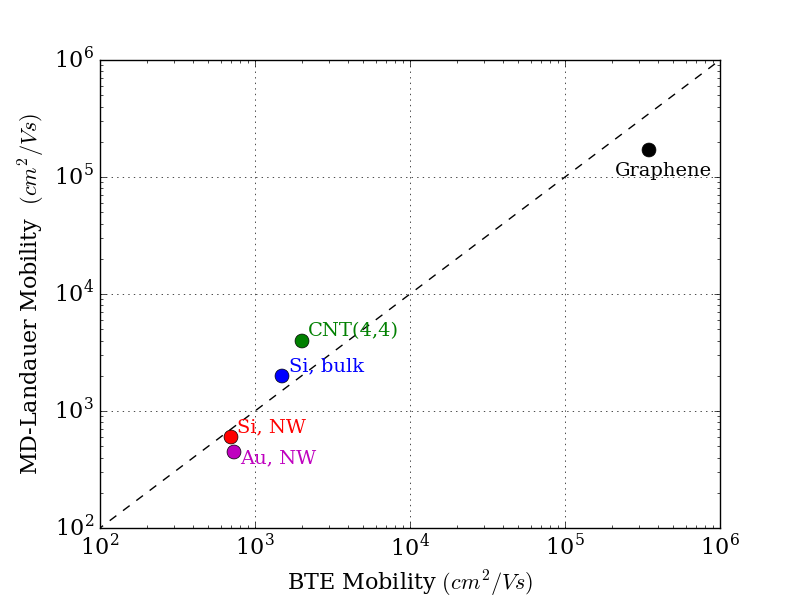}
\caption{Comparison of room temperature (300 K) mobilities calculated with BTE and with the MD-Landauer approach.  \label{fig:all_mobilities}}
\end{figure}

From the results presented above it is evident that the BTE and the MD-Landauer method  provide similar estimates for the phonon limited mobility in a variety of materials. To further illustrate this, we compare in Fig. \ref{fig:all_mobilities} room temperature mobilities for a number of systems (graphene, (4,4)-CNT, bulk silicon, silcon nanowire, and a gold nanowire). Some of the calculations have been detailed above, while the other are obtained in similar ways. All the calculations have been performed with DFT describing the electron transmission function. The figure illustrates that although the two calculation methods show deviations on quantitative level within a factor of 2-3, they generally predict the same mobility trends over more than two orders of magnitude. Both methods have their advantages and limitations. We will now discuss and compare various aspects of the two methods as they are implemented in ATK.

The BTE is the most rigorous and theoretically well founded of the two methods, but it relies on several assumptions: (i) It is clearly an assumption that the Boltzmann transport equation is an adequate description. This implies that any quantum interference effects are neglected, and any renormalization of the electronic band energies or eigenstates are not included. The effect of lowering the band gap in semiconductors at increasing temperatures is thus not included in the BTE approach, while the bandgap reduction is included in the MD-Landauer method.\cite{Franceschetti}
(ii) It is further assumed that the EPC can be described with first order perturbation theory through Fermi's golden rule. All scattering processes include a single phonon. Also, the construction of the perturbed Hamiltonian assumes that the screening is linear such that change in Hamiltonian from a sum of single atom displacements is the same as the change in Hamiltonian from the summed displacements. This has known limitations for e.g. polar materials where the long-wavelength Fr\"{o}hlich interaction is not correctly included.\cite{Bernardi2016} Note, that this is not a limitation in the BTE approach, but rather the way we calculate the electron-phonon coupling from finite displacements of individual atoms. (iii) The phonons are assumed to be described within the harmonic approximation. This means that anharmonic phonon-phonon couplings are not included. With these assumptions the BTE rigorously describes the scattering processes taking correctly into account the finite energy difference from initial to final electron states under absorption or emission of a phonon. The phonon occupation is also correctly described by a Bose-Einstein distribution.

The MD-Landauer method is not as well theoretically founded as the BTE. However, from the Born-Oppenheimer approximation we can argue that the electronic motion is much faster than the nuclear ones. If we are considering a finite and short MD region, an incoming electron passing through the MD region, essentially experiences a fixed potential landscape setup by the atoms in their instantaneous positions. Nevertheless, the MD-Landauer method does not correctly include finite energy transfer between the electronic and phonon systems, and assisted processes, where e.g. an electron absorbs a phonon to reach a higher lying final state, are not included. An advantage with the MD-Landauer approach is that it is not limited to first order perturbation theory. Given the perturbation in the Hamiltonian that is caused by the displaced atoms, the Green's function is solved exactly. Also, contrary to the way we calculate the EPC for the BTE approach, the MD-Landauer method does not assume the linear screening discussed above, and long wavelength Fr\"{o}hlich scattering will thus be included, provided the length of the MD region is long enough. The MD simulations naturally include anharmonic effects, which might be important at temperatures above the Debye temperature for the respective material. In the low temperature limit, the MD simulations will on the other hand not be correct since the phonon modes are occupied according to a Boltzmann distribution rather than the Bose-Einstein distribution implying that zero-point motions are not included.

In terms of applications, the two approaches also have different advantages and shortcomings. For bulk materials the BTE requires that both $\mathbf{k}$ and $\mathbf{q}$ are sampled on fine grids, resulting in a six-dimensional sampling, which is demanding computationally as well as memory wise. In the MD-Landauer approach one should only converge the transverse $\mathbf{k}$-point sampling for the transmission calculation in bulk systems. In addition to this, the cross sectional size of the unit cell must be converged. One also need to converge the sample averaging over different MD snapshots.

A clear advantage of the BTE approach is that it is relatively easy to obtain mobilities at many different temperatures. For the BTE calculations, the most time-consuming part is the calculations of the electron-phonon coupling matrix elements, which are temperature independent. The subsequent mobility calculations for different temperatures are computationally relatively inexpensive. For the MD-Landauer method, all calculations need to be redone for every temperature.

For device calculations, inclusion of EPC leads to a very substantial complication of the calculations, when treated with NEGF\cite{FrederiksenPRB2007,LuisierPRB2009,LuisierPRB2014}. The MD-Landauer method could, on the other hand, be included in device calculations without significant extra computational load. This potential application of the MD-Landauer approach will be pursued in future works.

Finally, the MD-Landauer approach is also applicable for studying EPC in amorphous systems, for which the BTE cannot be used since the electronic band structure is not well defined. Inclusion of other scattering mechanisms such as defect scattering or grain-boundary scattering is likewise relatively easy to include in the MD-Landauer approach.

\section{Conclusion\label{sec:conclusion}}
We have introduced a conceptually simple approach based on molecular dynamics (MD) and the Landauer transmission for calculating phonon-limited mobilities and resistivities. The results obtained with the MD-Landauer method are compared with values obtained from the Boltzmann transport equation (BTE). For several one-dimensional as well as bulk systems the results from the two methods are in good agreement with each other as well as with available experimental results. Our first-principles calculations further support the conclusion of enhanced electron-phonon coupling in nanowires previously indicated by tight-binding simulations. Advantages and shortcomings of the two methods were discussed. The MD-Landauer approach is a memory-efficient and computationally appealing alternative with a predictive power at the level of state-of-the-art BTE solvers for studying EPC in bulk and nano-scale systems.

\begin{acknowledgments}
The authors acknowledges support from Innovation Fund Denmark, grant Nano-Scale Design Tools for the Semiconductor Industry (j.nr. 79-2013-1) and from the European Commission’s Seventh Framework Programme
(FP7/2007–2013), Grant Agreement III–V-MOS Project No. 619326.
The Center for Nanostructured Graphene (CNG) is sponsored by the Danish Research Foundation, Project DNRF58.
\end{acknowledgments}


\end{document}